\documentclass[sn-mathphys-num]{sn-jnl}

\usepackage{graphicx}%
\usepackage{multirow}%
\usepackage{amsmath,amssymb,amsfonts}%
\usepackage{amsthm}%
\usepackage[title]{appendix}%
\usepackage{xcolor}%
\usepackage{textcomp}%
\usepackage{manyfoot}%
\usepackage{booktabs}%
\usepackage{algorithm}%
\usepackage{algorithmicx}%
\usepackage{algpseudocode}%
\usepackage{listings}%
\usepackage{physics}
\usepackage[export]{adjustbox}

\begin{document}

\title[Article Title]{Quantum Optical Approach to the $K$ Nearest Neighbour Algorithm}


\author[1]{\fnm{Vivek} \sur{Mehta}}\email{vivek\_1921ph05@iitp.ac.in}

\author[2,3]{\fnm{ Francesco} \sur{Petruccione}}\email{petrruccione@sun.ac.za}

\author*[1]{\fnm{Utpal} \sur{Roy}}\email{uroy@iitp.ac.in}

\affil*[1]{\orgdiv{Department of Physics}, \orgname{Indian Institute of Technology Patna}, \orgaddress{\street{Bihta}, \city{Patna}, \postcode{801103}, \state{Bihar}, \country{India}}}
\affil[2]{\orgdiv{Department of Physics, School of Data Science and Computational Thinking}, \orgname{Stellenbosch University}, \orgaddress{\postcode{7604}, \state{Stellenbosch}, \country{South Africa}}}

\affil[3]{ \orgname{National Institute for Theoretical and Computational Sciences (NITheCS)}, \orgaddress{ \postcode{7604}, \state{Stellenbosch}, \country{South Africa}}}


\abstract{We construct a hybrid quantum-classical approach for the $K$-Nearest Neighbour algorithm, where the information is embedded in a phase-distributed multimode coherent state with the assistance of a single photon. The task of finding the closeness between the data points is delivered by the quantum optical computer, while the sorting and class assignment are performed by a classical computer. We provide the quantum optical architecture corresponding to our algorithm. The subordinate optical network is validated by numerical simulation. We also optimize the computational resources of the algorithm in the context of space, energy requirements and gate complexity. Applications are presented for diverse and well-known public benchmarks and synthesized data sets.}




\maketitle
\section{Introduction}
Applications of machine learning are rapidly emerging in different aspects of life, such as medical, security, online-based shopping, and services \textit{etc}. The main hurdle for implementation is the overshooting of computational resources towards the training and validation of a machine learning model. Quantum computation can play a vital role in optimizing such complexity by exploiting quantum resources and bringing in quantum algorithms, which paves the fascinating area, quantum machine learning \cite{rebentrost2014quantum, lloyd2014quantum, schuld2016prediction, blank2020quantum}. On the other hand, the noisy intermediate-scale quantum (NISQ) era impels the researchers to take up an alternative approach, where the synergy between classical and quantum computers is used to avoid the large quantum overhead. Thus, a hybrid quantum-classical approach is widely studied in the contexts of quantum eigensolver \cite{peruzzo2014variational,kandala2017hardware,jones2019variational}, optimization problems \cite{farhi2014quantum,wang2018quantum}, nonlinear differential equations \cite{lubasch2020variational,kyriienko2021solving}, and machine learning  \cite{mitarai2018quantum, schuld2020circuit, havlivcek2019supervised, benedetti2019parameterized}.

Hardware platforms, based on quantum optics formulation, have potential implications for information processing and computing tasks \cite{knill2001scheme, kok2007linear, braunstein2005quantum}, which are translated by using the \emph{Integrated programmable photonic chips} \cite{o2009photonic} to further improve the energy optimization and speedup in information processing. Such platforms are well-tested for various quantum technology tasks like quantum communication \cite{metcalf2014quantum, bouchard2022quantum}, quantum simulations \cite{aspuru2012photonic}, boson sampling \cite{tillmann2013experimental}, machine learning \cite{zhang2021optical} \textit{etc}.
The first hybrid algorithm used the reconfigurable photonic circuits as quantum processors \cite{peruzzo2014variational}. These algorithms exploit different quantum optical states like Fock state, squeezed vacuum state, coherent state \textit{etc.}, as primitive resources for information processing \cite{kruse2019detailed, wetzstein2020inference, omkarall}. A number of schemes exist, where coherent states are used for quantum communication \cite{xu2015experimental,kumar2017efficient,touchette2018practical}, quantum cryptography \cite{grosshans2002continuous,ma2018phase, lucamarini2018overcoming}, quantum parameter estimation \cite{joo2012quantum,zhang2013quantum,lee2020optimal}, quantum machine learning \cite{chatterjee2016generalized, schuld2019quantum, killoran2019continuous}, universal computation \cite{jeong2002efficient,ralph2003quantum,lund2008fault}, image similarity measurements \cite{mehta2023quantum} \textit{etc}.

Among all the machine learning algorithms, $K$ nearest neighbour (KNN) model is widely studied \cite{mitchell2007machine} and protocols for implementing this model over the quantum computer exist \cite{schuld2014quantum,wiebe2014quantum}, where the information is encoded in the transition probability amplitude of the entangled qubits. It is seen that, such a qubit-based KNN algorithm consumes high quantum resources and is bound by long coherence time, which makes the implementation challenging in the NISQ era.

We propose a quantum-classical hybrid approach for the implementation of the KNN algorithm and optimize the requirement of quantum resources. A quantum optical architecture, involving both linear and nonlinear optical elements, is provided for the quantum part of our algorithm by using two resources: a single photon and a coherent state for information embedding. The proposed circuit includes the tasks of information encoding, processing, and photon detection. The proposed algorithm is also feasible for running over programmable integrated photonic circuits. The quantum optical circuit, involving the Walsh-Hadamard multiport devices, is validated in a quantum simulator. The working of the algorithm is demonstrated for a variety of well-known public benchmark data sets with sufficient accuracy.

In the following section, we discuss the basic idea of the algorithm for solving a supervised classification task. An optical scheme is presented for the Walsh-Hadamard transform, that is extensively used in our algorithm. The scheme for the training stage of the algorithm is presented in Sec. \ref{sec3}, followed by the protocol for similarity measurements. Section \ref{sec4} deals with the computational complexity and errors. We also numerically validate a prototypical optical network using a numerical simulator in sub-Sec. \ref{sec5a}, and establish the functionality of the driven distance metric from our protocol by solving a classification task over some well-known public benchmark and synthesized data sets in sub-Sec. \ref{sec5b}. We summarize our work and point out possible implications in the concluding section.

\section{Pre-requisites for the Quantum Algorithm}
\label{sec2}
Before we frame the main quantum algorithm, it is necessary to point out salient concepts, which will be used in our subsequent analysis.

\noindent\textit{\textbf{$K$-Nearest-Neighbour Algorithm:}}
Our training dataset $\mathcal{D}=\{(\mathbf{x}^m,g^m);m=1,\dots, M\}$ consists of $M$ labeled order pairs, such that the first element represents a $N$ dimensional feature vector, $\mathbf{x}^m\in \mathbb{R}^N$, where $N=2^n$ and $n\geq 1$, whereas the second element designates the corresponding class, $g^m$ ($\in\mathcal{G}$). Here, $\mathcal{G} :=\{c_1,\dots,c_s\}$ is a class set of total $s$ number of classes. In the supervised machine learning task, given the value of an unseen feature vector $\tilde{\mathbf{x}}$, the class $\tilde{g}$ is predicted to have a high probability for its true class by learning a pattern over the labeled dataset. A KNN algorithm is used for predicting the class of an unseen data point \cite{mitchell2007machine}, and it is a non-parametric algorithm, where one doesn't need a learning function and optimizing learning parameters, unlike a parametric machine learning model. While computing the \textit{closeness} between the training and the test data using well-defined distance metrics, it makes small structural assumptions and makes the predictions quite accurate, but sensitive towards the training data errors like outliners and wrong class labels. The working of the KNN algorithm is described below:
\vspace{0.3cm}
\hrule
\vspace{0.1cm}
\noindent \textit{Training stage}: Add a training database $\mathcal{D}$ and a test data $\mathbf{\tilde{x}}$.\\
\noindent \textit{Classification stage}: \emph{Step 1}: Compute distances between $\mathcal{D}$ and a test data $\mathbf{\tilde{x}}$, and save into a set, $\{d(\mathbf{x}^m,\mathbf{\tilde{x}})\}_{m=1}^M$.\\
\emph{Step 2}: \emph{i)} Choose a hyperparameter $K$, and sort $K$ training data points, nearest to $\mathbf{\tilde{x}}$, to prepare a set, $\mathcal{P}=\{(\mathbf{x}^1,g^1),\dots,(\mathbf{x}^K,g^K)\}$; \emph{ii)} Make the class assignment of $\mathbf{\tilde{x}}$  based on the majority voting as follows:\\
\indent $\Rightarrow$ For each class set $c_j\subseteq \mathcal{P}$, we compute the corresponding class average weight, ${g_j}=\frac{1}{K}\sum_{(\mathbf{x^i},g^i)\in c_j} g^i$, and store their values in another set, $\{(g_j,c_j);j=1,\dots,s\}$.\\
\indent $\Rightarrow$ Assign the label of $\tilde{x}$ as $\tilde{g}\leftarrow c_j=max(\{(g_j,c_j);j=1,\dots,s\})$,
where the function $max()$ returns the class, corresponding to the maximum-valued element $g_j$ from the given set.
\hrule
\vspace{0.2cm}
\noindent \textit{\textbf{Optical Network for the Walsh-Hadamard Multiport Device:}}

\noindent For running our quantum optical protocol, we are required to use multiport devices involving the Walsh-Hadamard transform. We first see its generic transformation relation in the recursive form:
\begin{equation}
\label{eq2.2.1}
	H_T=\frac{1}{\sqrt{2}}\begin{pmatrix}
	H_{T-1}&H_{T-1}\\
	H_{T-1} &-H_{T-1}
	\end{pmatrix},
\end{equation}
where recursive index $T> 0$, with the identity, $H_0=1$. The optical circuit deals with the \textit{dynamical evolution} of bosonic operators, which need to be decomposed into single and two qmodes (quantum modes) \cite{reck1994experimental}. These single and two qmodes transformations are physically realized by optical elements, like beam splitter, phase shifter, mirror, \textit{etc}. A bosonic system of $T$ qmodes has operators, $\{\hat{a}_j,\hat{a}^\dagger_j\}_{j=1}^T$, which satisfy the bosonic commutation relation, $[\hat{a}_j,\hat{a}^\dagger_j]=1$, where $\hat{a}_j$ and $\hat{a}^\dagger_j$ refer to the annihilation and creation operators, respectively. One can write the transformations,
\begin{align}
	\hat{b}^\dagger_k&=\sum_{j=1}^Tu_{kj}\hat{a}^\dagger_j, \textrm{ and} \label{eq2.2.3}	\\
    \hat{a}^\dagger_j&=\sum_{j=1}^T u^*_{jk}\hat{b}^\dagger_k,\label{eq2.2.4}
\end{align}
to represent the \emph{forward} and the \emph{backward} transformations between a set of $T$ inputs, $\{\hat{a}^\dagger_1,\hat{a}^\dagger_2,\dots,\hat{a}^\dagger_T\}$, and a set of \emph{T} outputs, $\{\hat{b}^\dagger_1,\hat{b}^\dagger_2,\dots,\hat{b}^\dagger_T\}$, creation operators.

The design of a generic Walash-Hadamard transformation relies on two-mode or four ports balanced beam splitters \cite{chabaud2018optimal,kumar2021optimal}, where the forward transformation of a two modes lossless beam splitter \cite{campos1989quantum} is
\begin{equation}
\label{eq2.2.5}
	BS=e^{i\phi_0}\begin{pmatrix}
	cos\gamma e^{i\phi_\tau} & sin\gamma e^{i\phi_\rho} \\
	-sin\gamma e^{-i\phi_\rho} & cos\gamma e^{-i\phi_\tau}
	\end{pmatrix}.
\end{equation}
Here, $\tau=cos^2\gamma$, where $0\leq\gamma\leq\frac{\pi}{2}$, and  $\rho=(1-\tau)=sin^2\gamma$ are called transmittance and reflectance, respectively. Its associated phases, $\phi_0,\;\phi_\tau \textrm{ and } \phi_\rho$ are respectively called global phase, transmitted phase and reflected phase. For designing a balanced lossless beam splitter, namely the matrix, $H_1$, one can take $\phi_0=0$, $\gamma=\frac{\pi}{4}$, and $\phi_\tau=\phi_\rho=\frac{\pi}{2}$.
\begin{figure}[ht]
    \centering
	\includegraphics[width=8cm,height=4cm]{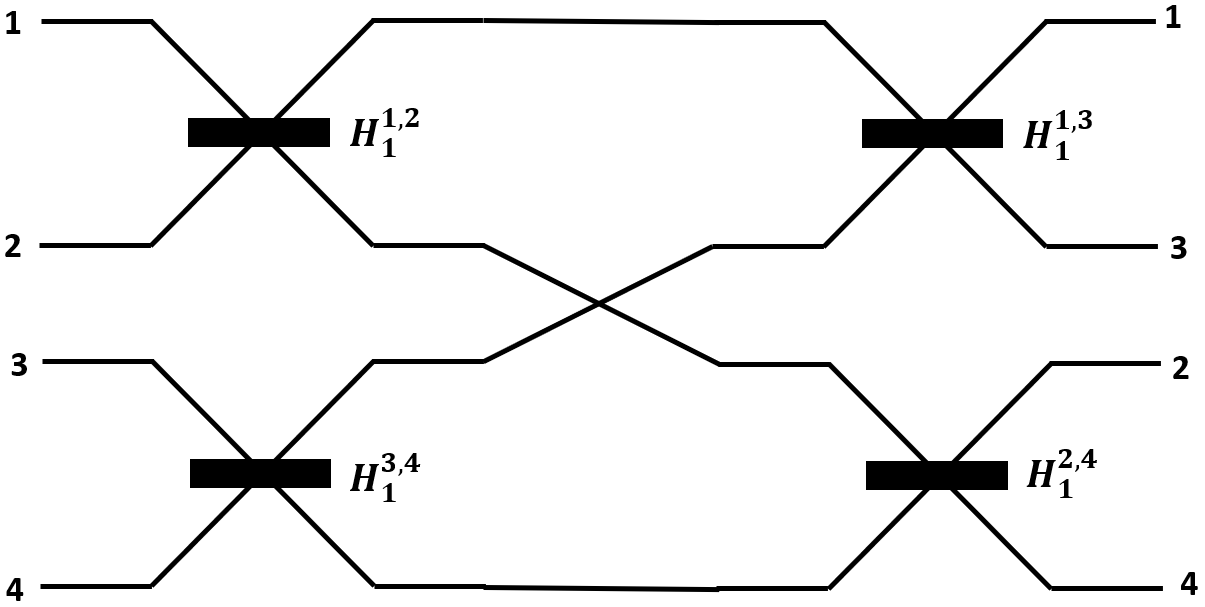 }
	\caption{Optical network architecture for realizing the Hadamard transformation, $H_2$, following Eq. \ref{H2}.}
    \label{fig1}
\end{figure}
The Walsh-Hadamard transformation for four qmodes is given by $H_2$, which is prepared by cascading $H_1$  as follows
\begin{align}
    H_2&=(H_1^{1,3}H_1^{2,4})(H_1^{1,2}H_1^{3,4})\notag\\
    &=\frac{1}{2}\begin{pmatrix}
    1&\textrm{ }1&\textrm{ }1&\textrm{ }1\\
    1&-1&\textrm{ }1&-1\\
    1&\textrm{ }1&-1&-1\\
    1&-1&-1&\textrm{ }1\\
    \end{pmatrix}.
    \label{H2}
\end{align}
We use this idea to describe the realization of $H_2$ by an optical network, as depicted in Fig. \ref{fig1}. It can be generalized for any $T$ by using $\frac{1}{2}Tlog_2T$ number of beam splitters instead \cite{chabaud2018optimal}.\\
 \\

\section{Quantum Subordinate of KNN }
\label{sec3}
Our algorithm uses coherent states for information encoding. The coherent state is the displaced vacuum state, $\hat{D}(\alpha)\ket{0}=\ket{\alpha}$, where  $\alpha \in \mathbb{C}$ is the coherent state amplitude, an absolute square of which is the average photon number. Since, the information will be encoded in the phase of a multimode coherent state, and hence it is required to map the data points within the bounded phase, $[0,\frac{\pi}{2}]$. Hence, we write the $k$-th feature of the $m$-th data point as
\begin{equation}
    \label{eq3.0}
    \theta_k^m=\frac{\pi}{2}\times\left[\frac{(x_k^m-x_k^{min)}}{(x_k^{max}-x_k^{min})}\right].
\end{equation}
Here, the term in the square bracket is out of the min-max scaling. Two terms, $x_k^{max}$ and $x_k^{min}$, are the smallest and largest values, respectively, for a particular feature ($k$-th) across all the data points. For describing the proposed quantum protocol to build a KNN scheme using multimode coherent state, we represent the training and test data points in terms of spatial qmodes as
\begin{align} \ket{\mathcal{D}}&=\frac{1}{\sqrt{M}}\sum_{m=1}^M\ket{1}_m\bigotimes_{k=1}^N\ket{e^{i\theta_k^m}\frac{\alpha}{\sqrt{N}}}_k \quad \textrm{and}	   \label{eq3.1}\\
    \ket{\mathbf{\tilde{x}}}&=\bigotimes_{k=1}^N\ket{e^{i\tilde{\theta}_k}\frac{\alpha}{\sqrt{N}}}_k,
\label{eq3.2}
\end{align}
where $\{\theta_k^m; k=1,\dots, N\}_{m=1}^M$ and $(\tilde{\theta}_k; k=1,\dots, N)$ represent the features of the training set and the test data in terms of phases, respectively.
$\ket{\mathcal{D}}$ consists of two multimode states, such that the first is the distribution of single-photon over $M$ qmodes (equal to the training data points), while the second is the $N$-qmodes (equal to the number of features) coherent state, with $m$ and $k$, being the entries of data points from the database and their features, respectively. The process of classical information encoding into the physical state of a quantum system may be referred to as the \textit{training stage} of the KNN algorithm.
\begin{figure*}
    \includegraphics[width=13cm,height=6cm, left]{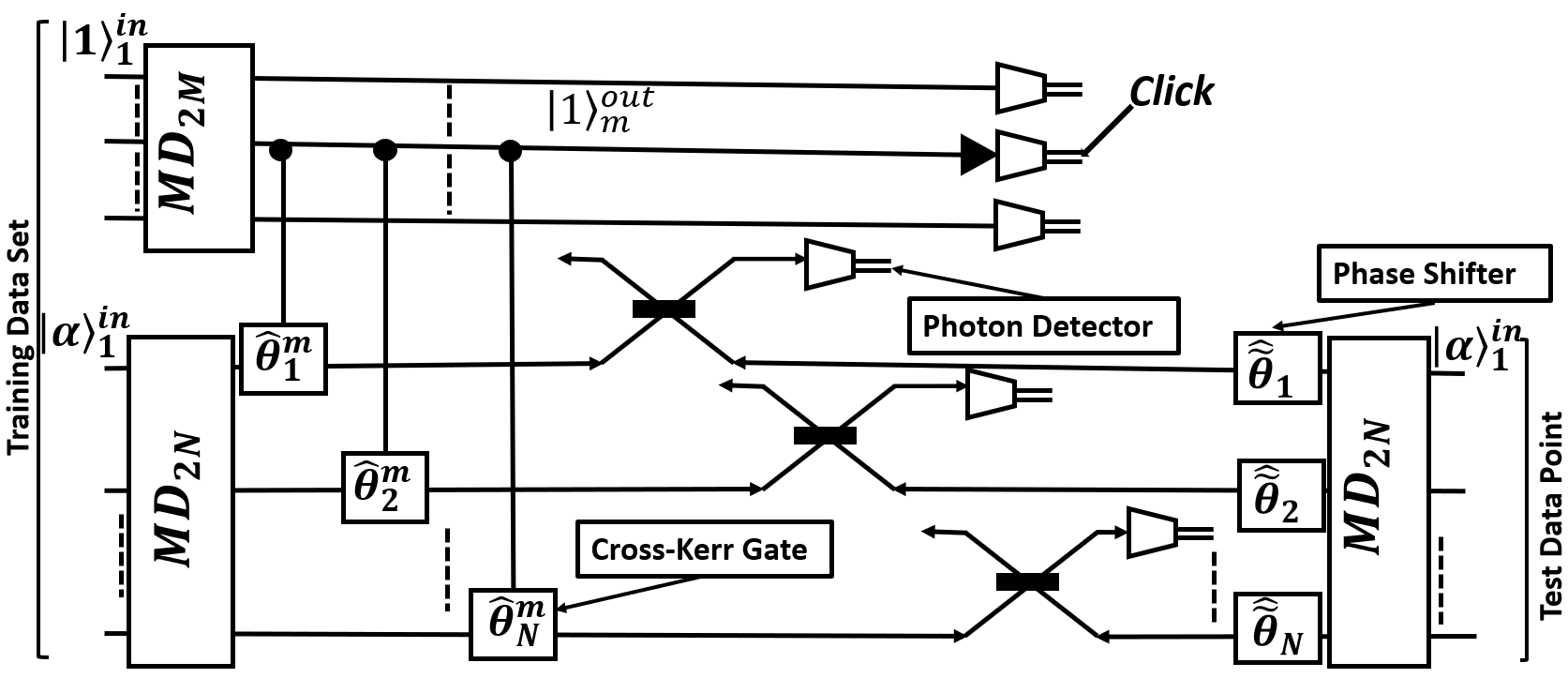}
    \caption{The optical scheme for the quantum subordinate of the KNN. The encoding of the training data set and the test data are shown on the left and the right edges of the diagram, respectively. The middle part presents the working of the protocol for obtaining the coherent distance metric by interfering with the phase-distributed coherent states, which are prepared with three multiport devices (MD) with subscripts, denoting the number of input of output ports.}
    \label{fig3}
\end{figure*}

Preparation of the state of Eq. (\ref{eq3.1}) can be explained through linear and nonlinear optical transformations. The distributed single-photon multimode state, $\frac{1}{\sqrt{M}}\sum_{m=1}^M\ket{1}_m$, requires embedding a single photon in any input port of a multiport device. However, for simplicity, we put it into the first input port, leaving the resultant output state as
\begin{align}
\label{eq3.3}
\ket{1}_1\bigotimes_{l=2}^M\ket{0}_l^{in}&=\hat{a}^{\dagger}_1\bigotimes_{l=1}^M\ket{0}_l^{in} \notag \\
	&\rightarrow\sum_{m=1}^Mu_{1m}\hat{b}^\dagger_m\ket{0}_m^{out}.
\end{align}	
Here, we use backward transformation (Eq. (\ref{eq2.2.4})) and consider the elements, $u_{1m}=\frac{1}{\sqrt{M}} \forall\textrm{ } m$, to prepare the output state as a single photon in a uniform linear combination of $M$ qmodes,
\begin{equation}
\label{eq3.4}
    \ket{1}_1\bigotimes_{l=2}^M\ket{0}_l^{in}\rightarrow\frac{1}{\sqrt{M}}\sum_{m=1}^{M}\ket{1}_m^{out}.
\end{equation}
Similarly, a single resource coherent state when fed into the \emph{first} input port of the multiport device, the output state becomes
\begin{equation}
\label{eq3.8}
    \bigotimes_{k=1}^N\ket{\frac{\alpha}{\sqrt{N}}}_k^{out}.
\end{equation}
Generic transformation of the multimode coherent state by its linear optical circuit is discussed in the appendix of the reference \cite{mehta2023quantum}. A schematic is provided in Fig \ref{fig3}, where the single photon in a $m$-th output qmode is associated with the features of the $m-$th index training data point, encoded into the phases over the multimode coherent state by $N$ cross-Kerr gates  \cite{nielsen2002quantum, jeong2006quantum}. The output state is written as
\begin{equation}
    \label{eq3.9}
    \ket{1}_m\bigotimes_{k=1}^N\ket{e^{i\theta_k^m}\frac{\alpha}{\sqrt{N}}}_k.
\end{equation}
We require to distribute the phase of the coherent state between $[0,\frac{\pi}{2}]$. Though a nonlinear media in general has small nonlinearities, typically in the order of $\chi\approx 10^{-18}$ \cite{boyd2020nonlinear}, this can be enhanced by using the electromagnetically induced transparency, and consequently, a phase change upto $\pi$ can be introduced \cite{tiarks2016optical}. The knowledge of the index of the encoded training data point is inferred from the click of that photon detector, which is attached to the corresponding spatial qmode. Since there is no prior knowledge of the indices of the encoding training data points, our encoding scheme consists of an equal probability of encoding for any training data point without bias (Eq. (\ref{eq3.1})). Moreover, in each run a set of $N$ phase shifters, $\{\hat{\tilde{\theta_k}}=e^{i\tilde{\theta}_k\hat{n}_k}\}$, apply in parallel over another multimode coherent state encode test data point (Eq. (\ref{eq3.2})) as shown in Fig. \ref{fig3}. Since the phase of light is not an absolute measurable physical quantity, we need to have a \emph{reference phase} before preparing the information-encoded multimode coherent state.

In the next stage of the scheme, the phase-distributed multimode coherent states, representing the training data point and the test data points, interfere via $N$ parallel balanced beam splitters. The output state is obtained as
\begin{align}
\label{eq3.10}
&\bigotimes_{m=1}^M\ket{1}_m\bigotimes_{k'=1}^N\ket{(e^{i\theta^m_{k'}}+e^{i\tilde{\theta}_{k'}})\frac{\alpha}{\sqrt{2N}}}_{k'}\notag\\
&\indent\indent\indent\indent\indent\bigotimes_{k=1}^N\ket{(e^{i\theta^m_k}-e^{i\tilde{\theta}_k})\frac{\alpha}{\sqrt{2N}}}_{k}.
\end{align}
Note that, $\{\theta_{k'}\}$ and $\{\theta_k\}$ both are the sets of pre-processed data points, however with two different dummy indices to distinguish between the output states that are coming from different ports of the balanced beam splitters.

\begin{figure*}[t]
\label{fig3.1}
    \centering
    \includegraphics[width=13cm,height=6cm]{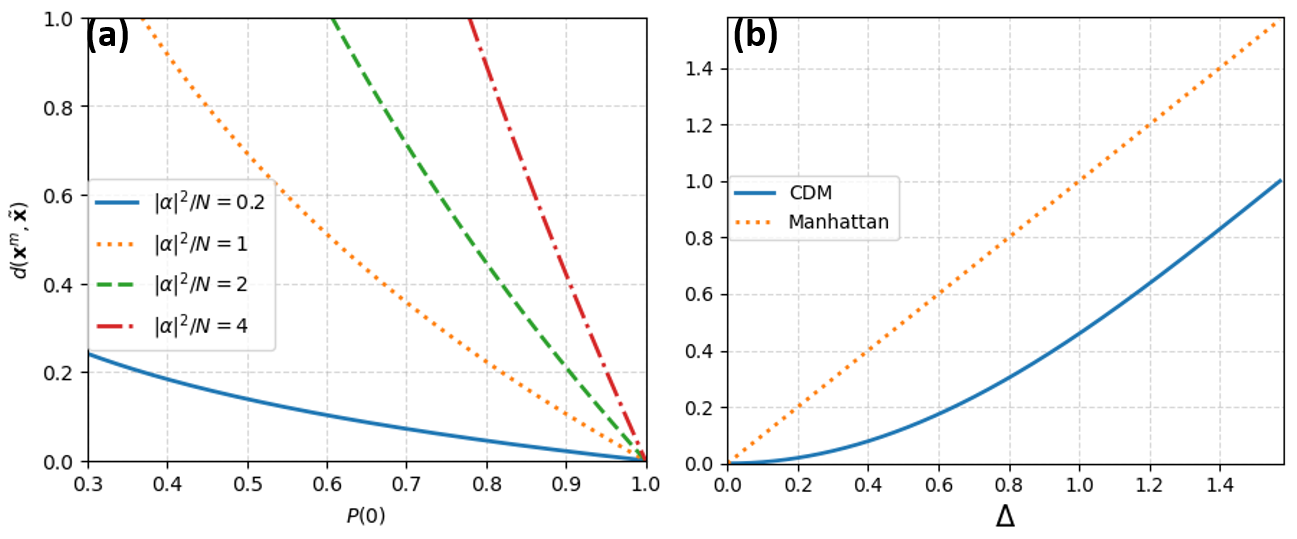}
    \caption{Plots for the distance metric with (a) the probability of projection to the no-photon subspace for various mean photon numbers, and (b) comparison of the Manhattan metric with the CDM \textit{w.r.t.} the variable between $0$ to $\frac{\pi}{2}$.}
    \label{fig5}
\end{figure*}

In the final stage of the circuit, we need to feed the multimode state, $\bigotimes_{k=1}^N\ket{(e^{i\theta^m_k}-e^{i\tilde{\theta}_k})\frac{\alpha}{\sqrt{2N}}}_{k}$, into $N$ parallel bucket photon detectors \cite{kok2010introduction}. Each bucket detector projects the state onto any two orthogonal subspaces, according to whether or not photons are present in the mode. We evaluate the probability of projecting into no photon subspace and obtain
\begin{align}
\label{eq3.13}
	P(0)&=\prod_{k=1}^N\left|\langle 0|(e^{i\theta^m_k}-	e^{i\tilde{\theta}_k})\frac{\alpha}{\sqrt{2N}}\rangle_{1}\right|^2 \notag\\
	&=\prod_{k=1}^N\left[exp\left\{\frac{-1}{2}\left|(e^{i\theta^m_k}-e^{i\tilde{\theta}_k})\frac{\alpha}{\sqrt{2N}}\right|^2\right\}\right]^2\notag\\
	&=exp\left\{-\frac{|\alpha|^2}{N}\sum_{k=1}^N\left(1-cos(\theta^m_k-\tilde{\theta}_k)\right)\right\}.
\end{align}
The above probability relation takes us to identify the distance metric, which can be referred to as \emph{coherent state} distance metric (CDM), in the form,
\begin{align}
\label{eq3.14}
	d(\mathbf{x}^m,\mathbf{\tilde{x}})&=\sum_{k=1}^N\left(1-cos(\theta^m_k-\tilde{\theta}_k)\right)\notag\\
	&=-\frac{N}{|\alpha|^2}ln(P(0)).
\end{align}
The above probability of projection on the no-photons subspace is bounded between $exp(-|\alpha|^2)$ and $1$. These bounds are driven by two extreme cases of the measurements: (i) both the data points are most dissimilar, i.e., $|\theta_k^m-\Tilde{\theta}_k|=\frac{\pi}{2} \text{ } \forall \text{ k}$, and (ii) both the data points are similar, \emph{i.e.},  $|\theta_k^m-\Tilde{\theta}_k|=0 \text{ }\forall \text{ k}$.

From Figs. \ref{fig5}(a) and \ref{fig5a}, the \emph{optimal} mean photon number for the resource coherent state primarily depends on the \emph{sensitivity} (distance/probability). Increasing the amplitude of the coherent state enhances the sensitivity, but diminishes the chance of noiseless propagation, as seen in the fidelity \textit{vs} transmissivity plot, (\ref{fig5a}). Both of the factors suggest an optimal mean photon number, given by $|\frac{ \alpha}{\sqrt{N}}|^2\approx 1$, implying $|\alpha|^2\approx N$. Figure \ref{fig5}(b) suggests the behavior of the CDM as the well-known distance metric called the Manhattan distance metric, where the \emph{phase parity}, $\Delta=|\theta_k-\Tilde{\theta}_k|$, takes values between $\left[0,\frac{\pi}{2}\right]$ for a given learning feature, excepting a nonlinearity in the CDM. Manhattan distance between two $N$-dimensional vectors is defined as $\sum_{k=1}^N|\theta_k-\tilde{\theta}_k|$, with regards to our learning feature space. It is worth mentioning that, the nonlinearity doesn't affect the capacity of the KNN algorithm for drawing the decision function when the inputs' feature are bounded.

Let us look at the \emph{statistical} error (discussion about the experimental error in the next Sec. (\ref{sec4})) for $\mathcal{O}(M)$ runs. Each qmode $k$ carries the similarity value $\ket{(e^{i\theta^m_k}-e^{i\tilde{\theta}_k})\frac{\alpha}{\sqrt{2N}}}_{k}$, which is \emph{statistically independent} from the rest qmodes, since information-carrying coherent states are product states. To avoid excessive quantum overhead, our estimator precision follows the shot-noise limit. Consequently, the error in the estimation of the true value of similarity by an unbiased estimator is inversely proportional to $\sqrt{\mathcal{O}(M)}$. An ultimate precision in similarity will not compromise on the quantum resource overhead and it will reach the Heisenberg limit \cite{lang2013optimal}. These computed distances will be stored in a classical co-processor with the corresponding labeled indices, which will be subsequently used in the classification stage of the algorithm.

\section{Computational Complexity of our Algorithm and Estimation of Possible Experimental Error}
\label{sec4}
\begin{figure}[ht]
    \centering
    \includegraphics[width=7cm,height=6cm]{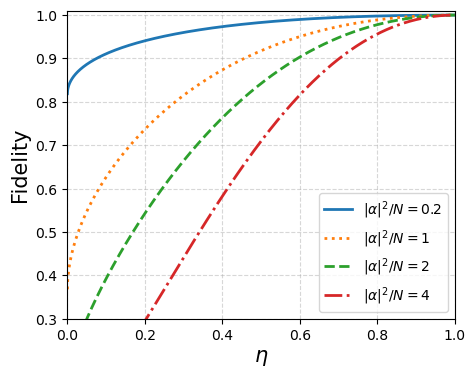}
    \caption{The variation of the fidelity with respect to the transmissivity, $\eta$, for different transmitting coherent state amplitudes through a nonlinear optical medium, where $\eta$ captures the influence of the propagating length and the medium characteristic.}
    \label{fig5a}
\end{figure}
\label{subsec3.1}
\begin{figure*}
\centering
    \includegraphics[width=13cm, height=12cm]{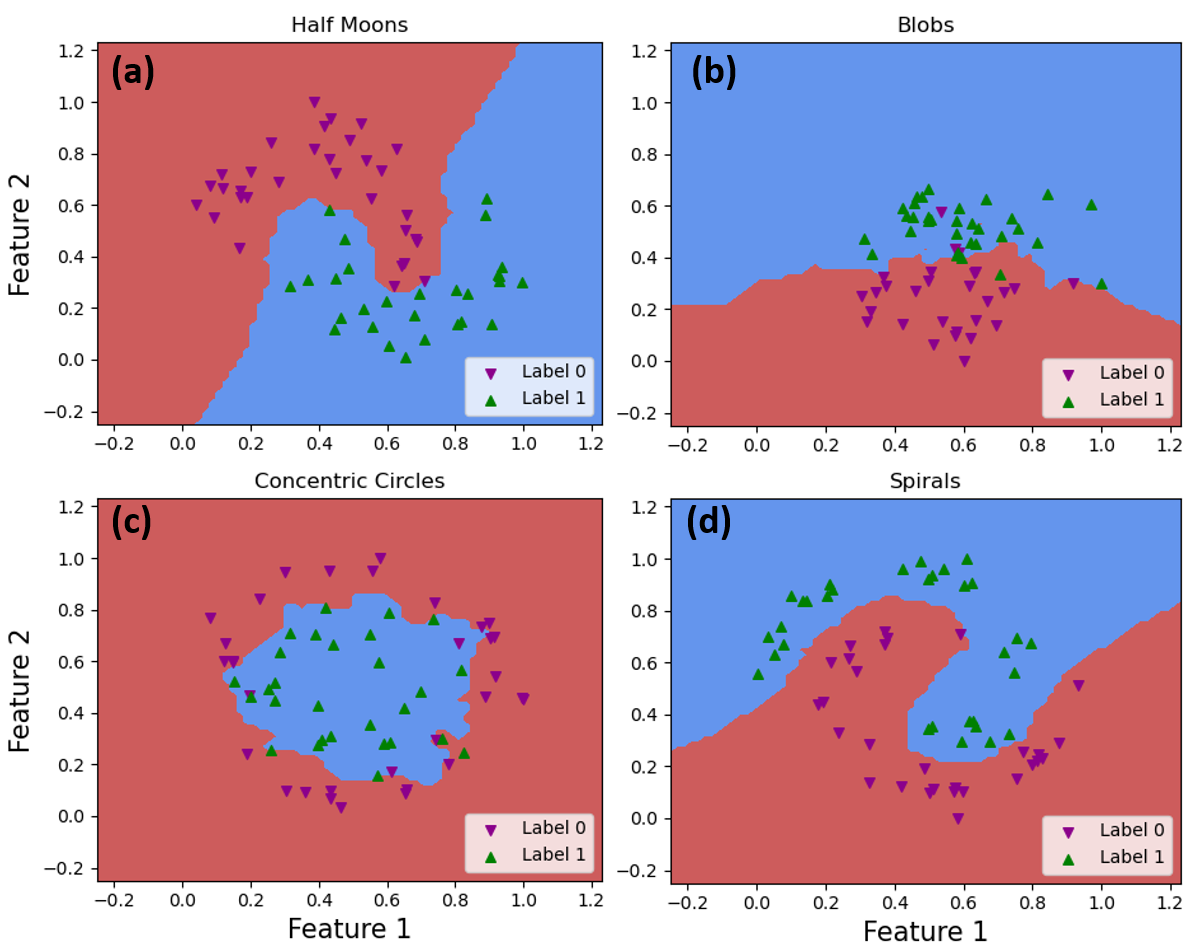}
    \caption{(Colour online) Decision boundary plots for random binary-labeled data sets. Each data set has  $200$ data points. Each data point has two learning features called features $1$ and $2$ and its label is an element of $\{0,1\}$. These decision plots for different synthesized data sets whose names are described above the plots. Here, we only show test instances that are represented by inverted triangles and upward triangles, along with the decision boundary.  }
    \label{fig7}
\end{figure*}
\begin{figure*}
\centering
	\includegraphics[width=13cm,height=6cm]{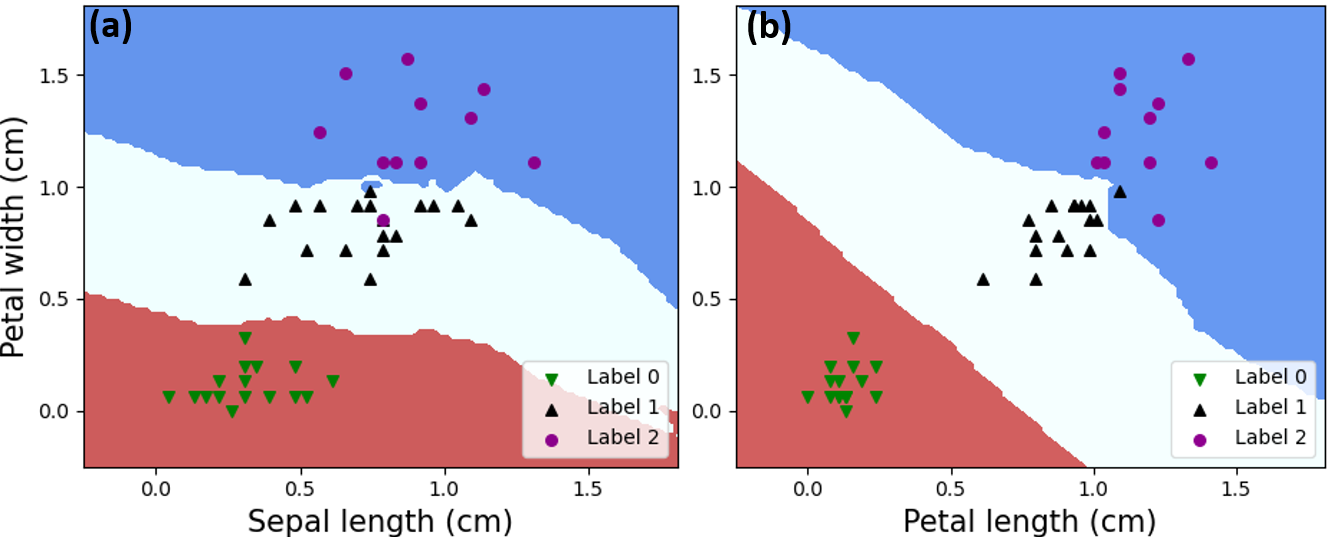}
	\caption{(Colour online) Decision boundary plots for (a) the data involving  Sepal length and Petal width, and (b) the data for Petal length and Petal Width. The class labels are designated in the plot legends as $0$, $1$, and $2$.  All lengths are represented in unit of centimeters. Here, we only show test instances.}
    \label{fig6}
\end{figure*}

The underlying algorithm involves \emph{quantumness} through two different states of light: a single photon state and the information-carrying phase distributed multimode non-orthogonal coherent states. The associated quantum computation faces unavoidable complexities. Such quantum computational complexities are usually threefold: \textit{space complexity}, \textit{gate complexity}, and \textit{energy resource} \cite{nielsen2002quantum}. The space complexity of a state vector lives in $N$-dimensional Hilbert space in the qubit model and is defined as the number of qubits ($log_2N$) required to store this vector. In the quantum optics model, the same state vector is stored by a single photon distributed over $N$-spatial modes \cite{cerf1998optical}, and hence space complexity should be the same as earlier, besides the physical Hilbert space associated with this state is different. We can conclude that the measure of the space complexity is independent of the quantum computational models and depends only on the dimension of the \emph{workable} Hilbert space \cite{arrazola2014quantum}. The idea of equivalence space complexity can also be defined for coherent states \cite{arrazola2014quantum}. On the other hand, the gate complexity is understood in terms of the number of optical gates required to implement the algorithm. Lastly, the energy resource requirement for any optical protocol measures the number of photons that are used to solve computational tasks.

It is important to address the computational complexity corresponding to our algorithm. First, we talk about the space complexity of our algorithm, where three quantum registers are consumed. One quantum register is a single photon distributed over $M$ spatial qmodes, and its equivalent space complexity is $log_2M$. The remaining two quantum registers are phase-distributed multimode coherent states.
We calculate the space complexity for the multimode coherent state following the method, developed by Arrazola  \emph{et al.} \cite{arrazola2014quantum}. The equivalent space complexity is found twice of $\mathcal{O}(N log_2N)$. We conclude that the space complexity decreases exponentially for the indices, representing the entry of the training data point, whereas the space complexity experiences only a polynomial decrease with the number of the learning features of each data point.

Regarding the gate complexity, the number of necessary elementary gates for our algorithm is discussed below. We use three Walsh-Hadamard multiport devices, two are of dimension $N$ qmodes, and the third is of $M$ qmodes. Hence, the total number of balanced beam splitters required to prepare all three devices are $\frac{1}{2}(Mlog_2M)+(Nlog_2N)$. In addition, we require $MN$ cross-Kerr interactions, $N$ phase shifters, and another $N$  balanced beam splitters. Overall, the total count of necessary elementary gates is in polynomial order of the total number of training data points. The energy resources of the algorithm will be given by the $(2N+1)$ total number of photons, where two resource coherent states contribute $2N$ photons with a single photon in addition. Therefore, energy requirement also increases polynomially with the number of learning features of the data points, while keeping a constant single photon for indexing the training data set.

In addition to computational complexity, it is worth estimating the possible experimental errors in the context of the implementation of the algorithm. Here, we consider two possible errors: amplitude damping and the inefficiency of the photon detectors. Amplitude damping of propagating coherent state, $\ket{\frac{ \alpha}{\sqrt{N}}}$, occurs due to photon absorption. It can be modeled by the fictitious beam splitter, interacting with the vacuum state with appropriate transmissivity, $\eta=e^{-\lambda L}$, where $\lambda$ is the loss coefficient of the nonlinear lossy medium (fiber optics or optical waveguide) and $L$ is the total distance of propagation. $\eta$ is a real physical parameter for a lossy medium, lying between $0$ (when $\lambda$ has a very high value) and $1$ (when $\lambda$ has a very low value). Usually, $\lambda$  has a significantly low value for high-grade commercial fibers. The output state after transmission being $\ket{\sqrt{\eta}\frac{ \alpha}{\sqrt{N}}}$, the fidelity manifests a measure of the effect of the transmitting medium over the input state and reads as $|\langle\frac{ \alpha}{\sqrt{N}}|\sqrt{\eta}\frac{ \alpha}{\sqrt{N}}\rangle|^2$. From Fig. \ref{fig5a}, we get the idea that the photon loss of transmitting coherent state increases with increasing its amplitude. The amplitude damping is the Gaussian noise and therefore is mitigated only through non-Gaussian operations. In the context of the effect of an inefficient photon detector without considering the dark counts, we consider the detection efficiency of a photon detector as $\tau$. Hence, the probability of detecting no-photon for the given Fock state $\ket{n}$ will be $P(0|n)=(1-\tau)^n$. The consequence of the inefficient detector is the failure to detect a photon (photons) in any mode, say $k$: $\ket{\beta}=\ket{(e^{i\theta_k}-e^{\tilde{i\theta}_k})\frac{\alpha}{\sqrt{2N}}}$; even if there is (are) photon (s). Thus, the probability for no-photon detection becomes
\begin{align}
    P(error)&=\sum_{n=1}^\infty |\langle n|\beta\rangle|^2(1-\tau)^n\notag \\
    &=\sum_{n=1}^\infty\Bigg[exp\left(-\frac{|\alpha|^2}{N}(1-cos(\Delta))\right) \notag\\
    & \hspace{1cm}\frac{|\alpha|^2}{N(n!)}(1-cos(\Delta))\Bigg](1-\tau)^n, \label{eq3.1.1}
\end{align}
where, $\Delta=|\theta_k-\Tilde{\theta}_k|$ is the parity between the phases. For avalanche photon detectors, the detection efficiency is about $90\%$. When we take the extreme case, $\Delta=\frac{\pi}{2}$, and fix the cutoff Fock state $n=5$, the net probability of failure for a given $|\alpha|^2=1$  and $N=2$ is evaluated to obtain $P(error)\approx3.2\%$.

\section{Validation of the Algorithm}

\subsection{Validation of Optical Architecture}
\label{sec5a}
\begin{table}[h]
\centering
\caption{Accuracies are described in row II  of all four random binary-labeled data sets. These data sets are balanced, i.e., the number of samples belonging to each class remains the same. Hyperparameter, $K=3$, the number of neighbors selected for assigning the class of test data.}
 \begin{tabular}{|c|c|c|c|c|}
\hline
Data set & Half moons & Blobs & Concentric Circles & Spirals \\\hline\hline
Accuracy(\%)& 98 & 90 & 88&100\\ \hline
\end{tabular}
\label{table2}
\end{table}
\begin{table*}[ht]
\centering
\caption{Verification of the algorithm by considering different public benchmark data sets: Wine, Sonar, and Iris Flower, as per column-I. The number of learning features and classes are described in columns, II and III, respectively. Column (IV) describes the type of sample data, belonging to each class, to see whether they are balanced or unbalanced. The best choice of hyperparameter is provided in column V, where the corresponding accuracy of classification over the given test data set is recorded in column VI. }
\begin{tabular}{|c|c|c|c|c|c|}
\hline
Data Set  & Number of & Number of & Sample & Hyperparameter& Accuracy \\
& Learning  & Classes &  & K & ( \%)\\
& Features & & & &\\ \hline\hline
Wine & 10 & 3 & Unbalanced & 20 & 84 \\ \hline
Sonar & 60 & 2 & Unbalanced & 5 & 80 \\ \hline
Iris Flower & 4 & 3 & Balanced & 3 & 98 \\ \hline
(Petal width,  & 2 & 3 & Balanced & 3 & 98 \\ \hline
Sepal length) &  &  &  &  &  \\ \hline
(Petal width,  & 2 & 3 & Balanced & 3 & 98 \\\hline
 Petal length) &  &  & &  & \\\hline
\end{tabular}
\label{table1}
\end{table*}
We check the working of the optical circuit through $H_2$ transformation by computing the probability of finding a single photon over the output ports while running it for several runs. Since, it is the Walsh-Hadamard transformation, the probability distribution is uniform for all output ports and also the same for a single photon injection from any input port. We have used the \emph{Strawberryfields} platform for checking \cite{killoran2019strawberry}, where an optical circuit, same as in Fig. \ref{fig1}, is implemented along with photon detectors at each of the output ports. We have four configurations, corresponding to each input port entry of a single photon. The said optical network is run several times in each configuration and the probability distribution is computed, which has come out as same for all the configurations. Simulation results verify the viability of the proposed optical architecture for $H_2$ and such numerical validation of optical architecture becomes necessary for its successful working in real devices. The same simulation can be generalized to Walsh-Hadamard multiport devices with different numbers of ports.

\subsection{Establishing the Validity of our Distance Metric for Classification Task}

\label{sec5b}
Now, we check the practicality of the obtained distance metric (Eq. (\ref{eq3.14})) for random binary-label data sets, which are generated from different functions after adding Gaussian random noises. Some well-known public benchmark data sets: Iris, Wine, and Sonar, are also used. We summarize the data sets, and their characteristics in Tab. \ref{table2} and \ref{table1}. We use \emph{Scikit-learn} Python's library to run the KNN algorithm, where our CDM is used as a distance metric. For testing purposes, we disintegrate the labeled data set into two, where $70\%$ data points are for training, while the remaining ones are for testing.  We observe that the considered data sets are quite nonlinear separable and diverse, even though the classification accuracies are high for both the synthesized data sets (Tab. (\ref{table1})) and the public benchmark data sets (Tab. (\ref{table2})). The decision boundary plots are demonstrated in Fig. (\ref{fig7}) and Fig. (\ref{fig6}) for random data sets and for data sets in Tab. \ref{table1}(rows 4 and 5), respectively. Notice that, we only demonstrate the decision boundary for such classification learning problems, which have only two learning features. A significant accuracy over the test data set emphasizes the practicality of the CDM as a good distance metric of our hybrid algorithm.

\section{Conclusion}
\label{sec6}
We have proposed a hybrid quantum-classical protocol, device by optical circuit, for supervised machine learning based on the KNN algorithm. Our proposed scheme uses a phase-distributed multimode coherent state to represent the data. We are usually skeptical about the use of phase of light, since it can't be directly measured, but can be estimated via other directly observed physical quantities. Besides this limitation, we can design a quantum optical algorithm that works with this approach of information encoding. We found that the computational complexity of our algorithm is efficient in both space and time. We also performed numerical simulations to validate our optical circuit as well as our proposed scheme by running a prototypical optical circuit over a photonic-based quantum computer simulator, which worked as desired. Finally, we have applied our distance metric to compute the closeness between the training and the test data sets of a few well-known public benchmarks and synthesized data sets. The outcome of the classification task establishes the utility of our distance metric and hence our proposed scheme.


\end{document}